\documentclass[%
 aip,
 amsmath,amssymb,
 reprint,%
]{revtex4-1}

\usepackage{graphicx}% Include figure files
\usepackage{dcolumn}% Align table columns on decimal point
\usepackage{bm}% bold math
\usepackage{hyperref}
\usepackage[utf8]{inputenc}
\usepackage[T1]{fontenc}
\usepackage{mathptmx}
\usepackage{etoolbox}
\usepackage{ulem}
\usepackage{setspace}%
\usepackage[multiple]{footmisc}
\usepackage[table]{xcolor}
\usepackage{multirow}
\usepackage{comment}
\usepackage{soul}

\relax
\newcommand{\BiSe}{Bi$_{2} $Se$_{3}$}
\newcommand{\BiTe}{Bi$_{2} $Te$_{3}$}

\newcommand{\Gbar}{\ensuremath{{\overline{\Gamma}}}}
\newcommand{\Kbar}{\ensuremath{{\overline{\text{K}}}}}

\newcommand{\teta}{$\theta_\mathrm{Sb}$}
\newcommand{\alfa}{$\alpha$}

\makeatletter
\def\@email#1#2{%
 \endgroup
 \patchcmd{\titleblock@produce}
  {\frontmatter@RRAPformat}
  {\frontmatter@RRAPformat{\produce@RRAP{*#1\href{mailto:#2}{#2}}}\frontmatter@RRAPformat}
  {}{}
}%
\makeatother
\begin{document}

\preprint{AIP/123-QED}

\title{Mastering the growth of antimonene on \BiSe: strategies and insights}

\author{Roberto Flammini}
\affiliation{CNR-Istituto di Struttura della Materia (CNR-ISM), Via del Fosso del Cavaliere 100, 00133 Roma, Italy}
\email{Roberto.Flammini@cnr.it}
\author{Conor Hogan}
\affiliation{CNR-Istituto di Struttura della Materia (CNR-ISM), Via del Fosso del Cavaliere 100, 00133 Roma, Italy}%
\affiliation{Dipartimento di Fisica, Universit{\`a} di Roma ``Tor Vergata'', Via della Ricerca Scientifica 1, 00133 Roma, Italy}
\author{Stefano Colonna}
\author{Fabio Ronci}
\affiliation{CNR-Istituto di Struttura della Materia (CNR-ISM), Via del Fosso del Cavaliere 100, 00133 Roma, Italy}%
\author{Mauro Satta}
\affiliation{CNR-Istituto per lo Studio dei Materiali Nanostrutturati (CNR-ISMN), Sapienza University of Rome, P. le Aldo Moro 5, 00185 Roma, Italy}
\author{Marco Papagno}
\affiliation{Dipartimento di Fisica, Universit\`{a} della Calabria, Via P.Bucci, 87036 Arcavacata di Rende (CS), Italy}
\affiliation{Laboratorio di Spettroscopia Avanzata Dei Materiali, STAR IR, Via Tito Flavio, Universit\`{a} Della Calabria, 87036, Rende (CS), Italy}
\author{Ziya S. Aliev}
\affiliation{Baku State University, AZ1148 Baku, Azerbaijan}
\author{Sergey V. Eremeev}
\affiliation{Institute of Strength Physics and Materials Science, Russian Academy of Sciences, 634055 Tomsk, Russia}
\author{Evgueni V. Chulkov}
\affiliation{Saint Petersburg State University, 198504 Saint Petersburg, Russia}
\affiliation{Donostia International Physics Center (DIPC), P. de Manuel Lardizabal 4, 20018 San Sebasti\'{a}n, Basque Country, Spain}
\author{Zipporah R. Benher}
\affiliation{University of Nova Gorica, Vipavska 11c, 5270 Ajdovscina, Slovenia}
\affiliation{CNR-Istituto di Struttura della Materia (CNR-ISM), Strada Statale 14 km 163.5, 34149 Trieste, Italy}
\author{Sandra Gardonio}
\affiliation{University of Nova Gorica, Vipavska 11c, 5270 Ajdovscina, Slovenia}
\author{Luca Petaccia}
\author{Giovanni Di Santo}
\affiliation{Elettra Sincrotrone Trieste, Strada Statale 14 km 163.5, 34149 Trieste, Italy}
\author{Carlo Carbone}
\author{Paolo Moras}
\author{Polina M. Sheverdyaeva}
\affiliation{CNR-Istituto di Struttura della Materia (CNR-ISM), Strada Statale 14 km 163.5, 34149 Trieste, Italy}

\date{\today}% It is always \today, today,
\begin{abstract}

Antimonene, the two--dimensional phase of antimony, appears in two distinct allotropes when epitaxially grown on Bi$_2$Se$_3$: the puckered asymmetric washboard ($\alpha$) and buckled honeycomb ($\beta$) bilayer structures. As-deposited antimony films exhibit varying proportions of single $\alpha$ and $\beta$ structures. We identify the conditions necessary for ordered, pure-phase growth of single to triple $\beta$-antimonene bilayers. Additionally, we determine their electronic structure, work function, and characteristic core-level binding energies, offering an explanation for the relatively large chemical shifts observed among the different phases. This study not only establishes a protocol for achieving a single $\beta$ phase of antimonene but also provides key signatures for distinguishing between the different allotropes using standard spectroscopic and microscopic techniques.

\end{abstract}

\maketitle

%\{}
\section{\label{intro}Introduction}

The term ``antimonene" refers to a two-dimensional (2D) layer of antimony atoms \cite{Hogan_PRB_pnictogen}. The properties of this material stem from its genuinely 2D structure combined with antimony's strong spin-orbit coupling, which is essential for supporting non-trivial topology in its electronic structure \cite{HasanKane,Fu2007,Qi2011,Wang2017}. This is why antimonene has attracted significant attention; fundamental research in areas such as spintronics \cite{LeiPRB2022_spin,sheve2023}, photonics \cite{Song_2017_optics,Lu2017_optics} and thermoelectricity \cite{Ersan2019_rev,Xue_2021_review,Sharma2017_thermo}, among others, as well as applications in electrochemistry \cite{GibajaJCMA2019,ZHANG2022_CO2,Li2023_CO2}, sensing \cite{Xue2019_RNA,Mayorga2019_sensing,Wang2024_catalysis}, energy storage \cite{MARIAPPAN2020_energy, Martinez2018_energy}, and biomedicine \cite{Tao2019_biomed,Uniyal2022_opticsmedicine}, could greatly benefit from its study.

The experimentally realized atomic structures of antimonene are limited to the puckered asymmetric washboard and the buckled honeycomb geometries, known as the $\alpha$ and $\beta$ phases, respectively. These structures correspond to the 2D limits of the orthorhombic A17 (black phosphorus) and the rhombohedral A7 (blue phosphorus) bulk phases of antimony. Predicted alternative structures, such as $\delta$-- and $\gamma$--antimonene, have been found to be thermodynamically unstable from their phonon dispersion spectra, exhibiting imaginary vibrational modes \cite{WangAMI2015}, which explains their absence in the experimental results. Additionally, there have been reports in the literature of a flat honeycomb allotrope of antimonene on Ag(111) \cite{Shao_flat} and Cu(111) \cite{Zhu2019NL,Niu_AdvMat_annealing}, although these structures were later identified as alloys \cite{ZhangPingPRM2022}.

The electronic band structures of both the $\alpha$ and $\beta$ allotropes have been extensively studied with first-principles calculations \cite{ZhangPRB2012, AkturkPRB2015, ZhangAngChe2016,JinPRB2016}. $\alpha$--antimonene shows a quasi-direct bandgap of 1.43 eV at the $\Gamma$ point (the indirect one being 1.18 eV), while $\beta$--antimonene exhibits an indirect bandgap of 2.28 eV, as estimated by density functional theory (DFT) calculations \cite{ZhangAngChe2016}. Both phases are predicted to exhibit interesting topological phenomena, paving the way to time-reversal symmetry--protected electron transport \cite{Wang2019_Ange, LuPRB_2021_alfa, Ji_JPCC2022_alfa_beta,  Lu2022_unpinned}. In particular, the $\alpha$-antimonene phase was recently reported to be topologically non-trivial, belonging to the quantum spin Hall class with a high spin Chern number \cite{Bai_PRB2022,Wang_2024}.  

In turn, the $\beta$ allotrope shows a strain-dependent bandgap \cite{Kripalani2018} as well as a thickness-dependent electronic band structure, ranging from that of a semiconductor to that of a topological semimetal \cite{ZhangAngChe2016}. Moreover, a MOSFET made by $\beta$-antimonene was successfully simulated, showing promising I/V characteristics \cite{Pizzi2016_transistor,Chang2018_TFT,Yin2020_FETbeta} and very recently, also the $\alpha$ allotrope was proposed in a transistor able to match the requirements of high performance devices \cite{ZhangIEEE2022}.  

Mechanical or liquid-phase exfoliation from bulk antimony primarily results in multilayer flakes \cite{GibajaACIE2016, AresADMA2016}. Therefore, significant effort is devoted to developing new methodologies for isolating single layers, potentially enabling the scalable production of high-quality flakes \cite{Lucherelli_Sb_exfoliation}. Another effective way to synthesize antimonene flakes is using van der Waals epitaxy in a tube furnace \cite{JiNatComm2016}. However, physical vapor deposition technique is particularly promising for producing 2D material layers with controlled structures. This method provides precise thickness control, is conducted in an ultra-high vacuum to prevent contamination, and is solvent-free, which reduces environmental impact.  Currently, the synthesis of $\alpha$ and $\beta$ allotropes as single layers relies on Sb evaporation on suitable substrates. Specifically, epitaxial growth of antimonene, whether in the \alfa\ or $\beta$  phase, has been demonstrated on materials such as \BiTe\ \cite{TaoLei_JAP_2016,Taolei2019APL,YaoLi2022}, Bi$_2$Se$_2$Te \cite{Kim2016SR,KimACSnano2017}, Sb$_2$Te$_3$ \cite{su2021topological}, PdTe$_2$ \cite{WuADMA2017}, graphene \cite{XinSun2018,Fortin_NL_2020}, highly ordered pyrolytic graphite \cite{LeSterPRB_2019_moire_alfa}, SnSe \cite{Lu2022_unpinned,LuPRB_2021_alfa}, TiSe$_2$ \cite{Ji_JPCC2022_alfa_beta} MoS$_2$ \cite{LeSterPRB_2019_moire_alfa} and on metals like W(110) \cite{Jaroch2024,Drozdz2024} and Bi(111) \cite{Markl2DM2018,Nakamura2024}. 

Among all the substrates, a prototypical topological insulator like \BiSe\ is of particular interest, as its surface structure is nearly lattice-matched with that of $\beta$--antimonene while naturally facilitating the formation of the \alfa\ phase during the Sb deposition process \cite{Flammini_2018, HoganACSNano2019,Hogan_PRB_pnictogen}. Surprisingly, $\beta$--antimonene becomes metallic in contact with bismuth selenide, inheriting its topological properties, due to a proximity effect \cite{TaoLei_JAP_2016, JinPRB2016, Holtgrewe2020_SciRep, su2021topological}. Moreover, the topological surface state (TSS) of Bi$_2$Se$_3$ with a $\beta$-antimonene overlayer was also shown to have much stronger surface localization with respect to the TSS hosted by the pristine substrate, revealing a giant and tunable out-of-plane spin polarization \cite{sheve2023}, of paramount importance for spin-charge interconversion devices \cite{yin2019selective,yin2023extracting}. Similarly, an even stronger surface localization and a nontrivial spin texture can be expected for thicker $\beta$-antimonene films \cite{Holtgrewe2020_SciRep}. Therefore, leveraging these phenomena in real devices requires a comprehensive understanding and complete control of the growth process.

In this article, we present results from our previous investigations on the Sb/\BiSe\ system, supplemented by unpublished data. A comprehensive analysis of Sb epitaxially grown on Bi$_2$Se$_3$ is then presented as a function of coverage and temperature, using a combination of scanning tunneling microscopy (STM), angle--resolved photoemission spectroscopy (ARPES) and core--level photoemission spectroscopy, complemented by DFT and atomistic thermodynamics calculations. These high-resolution, surface-sensitive techniques are particularly suitable for characterizing epitaxial growth and (electronic) structure of adlayers under ultra high vacuum (UHV).
The ultimate aim of this research is to determine the optimal conditions for achieving a specific antimonene phase, thereby enabling full exploration of its electronic properties.

\section{Experimental}

A clean and ordered surface of \BiSe(0001) was obtained by exfoliating a bulk sample in UHV conditions. The structural ordering of the sample was confirmed by low energy electron diffraction and by STM. Antimony was sublimated from a temperature-controlled Knudsen cell. All depositions were performed with the sample at room temperature (RT). 
In this work, due to the layered atomic geometry of both allotropes, we refer to a single antimonene 'sheet' to mean a continuous bilayer, extended across most of the sample surface.
Coverage is reported in terms of a percentage $\theta_\mathrm{Sb}$ of full coverage (FC) of the $\beta$--antimonene phase, i.e. 2 Sb atoms per Bi$_2$Se$_3$ surface unit cell.
The temperature calibration of the sample was performed by exploiting the known surface phase transition undergone by the system consisting of more than 0.75 FC of Sb deposited on Ag(111), as detailed elsewhere \cite{HoganACSNano2019}.

STM images were recorded by using an Omicron LT-STM housed in a UHV vacuum chamber with base pressure below 1$\times$10$^{-10} $ mbar. The STM images were acquired at 80 K using a W tip cleaned by electron bombardment in UHV. The STM scanner was calibrated by measuring the clean \BiSe(0001) surface. Bias voltage is referred to the sample, hence positive (negative) bias corresponds to empty (filled) states. 

ARPES and core level spectroscopy experiments were performed at two complementary beamlines, namely the VUV-Photoemission and BaDElPh beamlines of the Elettra synchrotron in Trieste (Italy) at RT with 20 and 75 eV photon energies. The VUV-Photoemission beamline endstation is equipped with a Scienta R4000 electron spectrometer, while the BaDElPh beamline endstation features a Specs Phoibos 150 analyzer. The angle between the photon beam (horizontal polarization) and the analyzer is 45$^{\circ}$ for the VUV setup and 50$^{\circ}$ for the BaDElPh setup. The photoelectrons were collected within the light scattering plane. 

The Sb/Bi$_2$Se$_3$ interface was modelled using DFT as implemented within the quantum-ESPRESSO suite.\cite{QE} 
The PBE exchange-correlation functional\cite{PBE} was used along with the Grimme-D2 van der Waals coupling.\cite{Grimme-D2} %\comm{Which vdW functional was implemented?} 
A plane-wave and pseudopotential framework (ultrasoft flavour\cite{DALCORSO2014337}) was adopted (cutoff 45 Ry). Spin-orbit coupling was included throughout. The substrate was modelled using a Bi$_2$Se$_3$ slab containing six quintuple-layers and using the experimental surface lattice constant of 4.143 \AA. Sb layers were added to both sides of the slab, with at least 20 \AA\ of vacuum separating periodic replicas. For the $\beta$-antimonene phase a 15$ \times $15$ \times $1 k-point mesh was used. The $\alpha$-Sb phase was modelled using a (1$ \times $4) supercell, and a 15$ \times $5$ \times $1 mesh. During geometry optimizations the Sb atoms and outermost 4 atomic layers of Bi$_2$Se$_3$ were allowed to relax freely.
Obtained geometries were consistent with previous works.\cite{HoganACSNano2019,Flammini_2018,Hogan_PRB_pnictogen} To provide further support for our results we have performed additional VASP \cite{vasp1,vasp2} calculations for $\beta$-antimonene on Bi$_2$Se$_3$. These calculations show similar results for the band structure to our quantum-ESPRESSO and earlier DFT data. \cite{Holtgrewe2020_SciRep,Holtgreve2021_Materials}

Core level shifts (CLS) were evaluated by DFT calculations, again with ultrasoft PBE pseudopotentials. The Sb atoms with the core hole were associated with a pseudopotential generated with a hole in the \textit{4d} orbitals. Integration in the reciprocal space were performed using the $\Gamma$ point after preliminary test calculations on the CLS (error of the order of tenths of meV). The cell used in the calculation was fixed at dimensions 2$\times$4$\times$7, with 4 alternate atomic layers of Bi and Se. Test calculations with cell 2$\times$4$\times$10 showed that the CLS error is again of the order of tenths of meV. The common reference state to compute the XPS shift was chosen to be a free Sb atom with a \textit{4d} core hole, and fixed at about 10~\AA\ from the topmost layer of Sb.

\section{Results}
\begin{figure}[t] 
	\includegraphics[width =0.95\columnwidth]{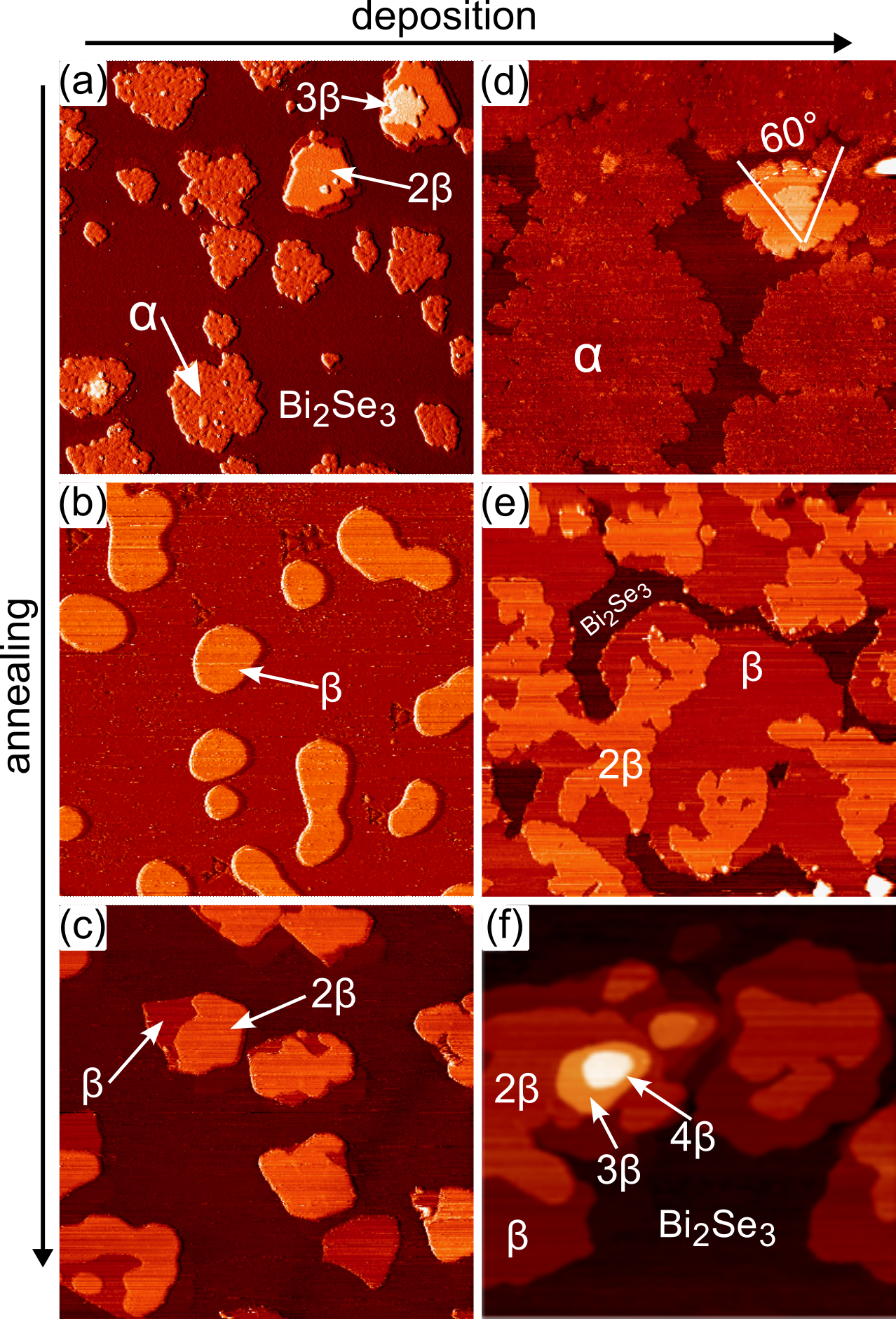}
	\caption{STM images taken at 80 K of the Bi$_2$Se$_3$ surface, \teta=0.5 coverage (a-c) of Sb
at RT (500$\times$500 nm$^{2}$, +1 V, 2 nA). (a) Prior to the annealing (b) after annealing to 473 K for 30 min; (c) after annealing to 473 K for 60 min. STM images taken at 80 K, for a Sb coverage of nearly \teta=1 at RT (d-f); (d) prior to the annealing (e) after annealing to 473 K for 45 min (f) 200$\times$200 nm$^{2}$ image (+1 V, 5 nA), after annealing at 473 K for 60 min. The labels $\alpha, \beta$, 2$\beta$, 3$\beta$ or 4$\beta$ stand for 1 BL $\alpha$, 1, 2, 3 or 4 BL $\beta$, respectively. The \BiSe\ substrate is depicted in the darkest colour.}  \label{Fig-STM}
\end{figure}

The typical appearance of the partially covered surface (\teta=0.5) at RT is displayed in Fig.~\ref{Fig-STM}(a). The islands appear jagged, the boundaries irregular. The $\alpha$-phase islands are characterized by randomly oriented stripes, whereas the $\beta$-phase islands appear flatter and more ordered. The structure of ordered and disordered domains and their stability were already addressed in Refs. [\onlinecite{Flammini_2018}] and [\onlinecite{HoganACSNano2019}], on the bases of atom-resolution STM, high energy resolution ARPES and DFT simulations.

Most of the islands show the $\alpha$-antimonene phase of 1 BL thickness, while some other flat islands of double and triple BL thickness, exhibit the $\beta$-antimonene atomic structure (lighter orange colour). Importantly, we observed neither large single BL domains of $\beta$--antimonene, nor more--than--1 BL--thick domains of $\alpha$--antimonene. 

After initial annealing at 473 K (Fig.~\ref{Fig-STM}(b)), $\alpha$ islands disappear and transform into the 1 BL $\beta$ islands \cite{HoganACSNano2019}. With further annealing, the effects of dewetting become apparent: the 1 BL islands disappear in favour of mostly 2 BL thick ones with the $\beta$--antimonene structure (Fig.~\ref{Fig-STM}(c)).

Starting with a larger coverage (slightly below \teta=1) at RT (Fig.~\ref{Fig-STM}(d)), the domains of $\alpha$--antimonene cover most part of the surface and are accompanied by 2 and 3 BL thick flat islands exhibiting the $\beta$ structure. 
It is interesting to notice that the 3 BL thick yellow island, with the shape of a slice of cake or pie, displays an angle of exactly 60$^\circ$, is reminiscent of its honeycomb structure and the honeycomb structure of the layer below. Upon annealing at 473 K for 45 min (that is, a duration well above the completion of the $\alpha$ to $\beta$ phase transition \cite{HoganACSNano2019}), the surface appears like in Fig.~\ref{Fig-STM}(e), where the $\alpha$ phase is no longer present and the 1 BL $\beta$ domains are reduced in favour of 2 BL $\beta$ domains. Moreover, part of the surface begins to uncover, and with prolonged annealing (60 minutes), as shown in panel (f), the surface becomes further exposed (dewetted), revealing $\beta$ domains with a thickness of up to 4 BL. 

We do not observe $\alpha$ domains thicker than 1 BL for RT growth, and no $\alpha$ domains of any thickness for annealed samples. Both findings are in line with the peculiar character of the $\alpha$ phase of being stable depending on the temperature \cite{HoganACSNano2019}    and on the thickness, as previously reported for Ge(110)/graphene supported antimonene \cite{Fortin_NL_2020}. While the formation of large ordered 1 BL $\beta$--antimonene was not observed without annealing, 2 and 3 BL $\beta$ domains can form already at RT, although they do not dominate the thickness distribution. 

%%%%%%%%%%%%%%%% ARPES %%%%%%%%%%%%

\begin{figure*}[t]
	\includegraphics[width=1\textwidth]{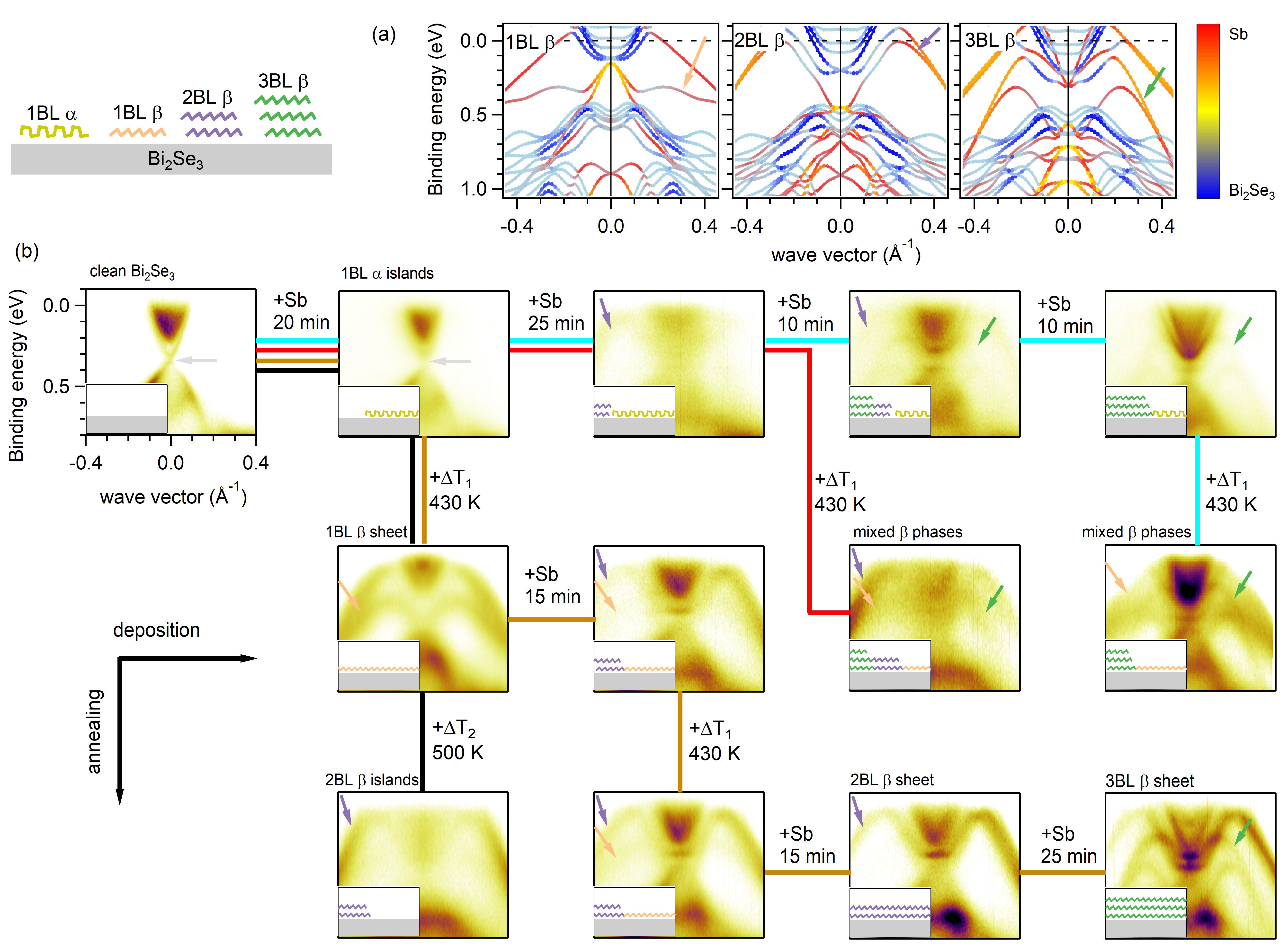}
	\caption{Panel (a) DFT band structure for 1, 2 and 3 BL of $\beta$-antimonene on Bi$_2$Se$_3$ along $\Kbar\Gbar\Kbar$ high symmetry direction. The characteristic electronic features are marked by colored arrows. 
(b) `Transit map' of the antimonene growth strategy. In analogy with a transit map (e.g. metro or Tube map), the coloured paths are referred to as cyan (deposition line), black (annealing line), red (mixed line) and brown ($\beta$ line). ARPES data along $\Kbar\Gbar\Kbar$ direction. Top left panel, schematics of the different antimonene phases. The insets schematically indicate the coverage and composition of the adlayers (see text for more details). The binding energy and wave vector scale are indicated on the first panel. The data are taken at RT with photon energy of 20 eV. } \label{Fig-ARPES}
\end{figure*}

What matters most in real devices, is the formation of a specific phase on a large scale. For this purpose, we studied the electronic band structures as a function of Sb coverage and annealing steps to identify each phase. We used the ARPES technique, whose typical field of view is about 100$\times$300$\mu$m$^{2} $. As a reference, the band structures displayed in Fig.~\ref{Fig-ARPES}(a) for 1, 2 and 3 BL $\beta$--antimonene on \BiSe\ were computed by DFT for comparison. 
We did not calculate the band structure of the $\alpha$ phase. 
Its incommensurate lattice (with respect to Bi$_2$Se$_3$) leads to several rotated domains \cite{HoganACSNano2019}: as a result, no clear spectroscopic features can appear in the ARPES data (see below). The DFT calculations for 1 and 2 BL $\beta$--antimonene are in line with the previous reports \cite{KimACSnano2017, Holtgrewe2020_SciRep}. The band structure of 3 BL $\beta$ resembles that of 2 BL, with an additional characteristic shape at the zone center, marked by a green arrow (top right corner). This feature, together with the other peculiar features of 1 BL and 2 BL structures (see arrows in the corresponding panels) will be used in the following as fingerprints of the different thicknesses in the ARPES spectra.

To give a more comprehensive overview of antimonene formation on \BiSe\ for different temperature and flux conditions, we present in Fig.~\ref{Fig-ARPES}(b) a collection of ARPES data in the form of a so-called 'transit map'. In order to reach a specific antimonene phase one needs to follow the four coloured lines that evidence the different preparation sequences discussed in the text. 
The ARPES data were acquired along the $\Kbar\Gbar\Kbar$ high symmetry direction. In the panel, the coverage increases from left to right while the annealing duration and temperature increases from top to bottom. The binding energy and wave vector scales are indicated on the top left map. Each ARPES panel is associated with a surface structure that is schematically represented in the inset in the bottom left corner. The colors of the arrows correspond to the different antimonene structures.

The topmost row, connected by the cyan path, shows the evolution of the band structure as a function of the coverage prior to annealing. The first panel shows the band structure of clean Bi$_2$Se$_3$ with its Dirac point marked by a grey arrow. The initial deposition of Sb leads to a weakening of the TSS of Bi$_2$Se$_3$ (grey arrow) without other clear spectroscopic features. This observation is in line with the poor structural order of the $\alpha$ allotrope and with the presence of different rotational domains \cite{HoganACSNano2019}. For a higher Sb coverage, we can see the appearance of weak bands related to the 2 BL $\beta$ allotrope, in agreement with the literature \cite{KimACSnano2017}. Together with the information provided by STM, we can conclude that epitaxial growth of the $\alpha$ allotrope is limited to 1 BL, while the Sb in excess is allocated in 2 BL $\beta$-islands. The high background can be attributed to the contribution of the dominating $\alpha$ phase, in agreement with the STM observation. Further Sb evaporation does not lead to an enhancement of 2 BL $\beta$ features, but to the emergence of new sharp states relative to the 3 BL thick $\beta$ phase (indicated by the green arrow). 

The black line follows the evolution of the surface prepared with a fixed coverage, mostly consisting of the $\alpha$ phase as confirmed by the STM measurements, followed by several annealing steps. When the surface is annealed for the first time, the bands of 1 BL $\beta$ clearly emerge (panel `1BL $\beta$ sheet'), in line with the literature \cite{HoganACSNano2019, Holtgrewe2020_SciRep, sheve2023}. Due to the higher surface atomic density of the $\alpha$ phase compared to the $\beta$ phase, full coverage in the $\beta$ phase is achieved with only two-thirds of the surface covered by antimonene in the $\alpha$ phase. Further annealing leads to the appearance of 2 BL $\beta$ islands, in line with STM findings, as shown in Fig.~\ref{Fig-STM}(c). This is also corroborated by the observation of clear fingerprints of the almost pure 2 BL $\beta$--antimonene phase emerging at the end of black line, i.e. after annealing at higher temperature. However, this procedure does not result in a complete 2 BL $\beta$ surface, due to the unavoidable presence of uncovered areas of the substrate, as shown in the STM data (see Fig.~\ref{Fig-STM}).  

To overcome the problem of the dewetting and to reduce the uncovered areas of the surface, it is natural to consider increasing the initial coverage of Sb before annealing. However, the annealing of thicker samples, that already contain 2 BL or 3 BL $\beta$ domains, leads to mixed phases as shown at the end of red/cyan paths: there is always a contribution of the 1 BL $\beta$ phase, as a result of the transformation from the 1 BL $\alpha$ phase. We can also observe the formation of the 3 BL $\beta$ phase, well distinguishable from its characteristic band structure. The higher the Sb coverage, the higher the contribution of the 3 BL $\beta$ phase, so that the 2 BL does not form as a single $\beta$ phase, in agreement with STM results. 

Very good quality 2 BL $\beta$ bands emerge only when Sb is deposited on top of a 1 BL $\beta$ phase, even without annealing (follow the brown line, second row). Moreover, further annealing does not undermine the quality of the surface and only the ulterior deposit of Sb improves the band structure (panel marked `2BL $\beta$ sheet'). The large--scale, full--coverage and single--phase 2 BL $\beta$--antimonene necessarily requires a template layer of 1 BL $\beta$. Similarly, the 3 BL $\beta$ phase can form on top of the 2 BL $\beta$ structure even without annealing (end of brown line, panel marked `3BL $\beta$ sheet'). Therefore, the ordered growth of 2 BL and 3 BL $\beta$ structures requires a three--step procedure (that is, growth+annealing+growth) across the 1 BL $\alpha$ and 1 BL $\beta$ phases.

%%%%% THERMODYNAMICS %%%%%%

\begin{figure}[tb!] 
	 \includegraphics[width =0.99\columnwidth]{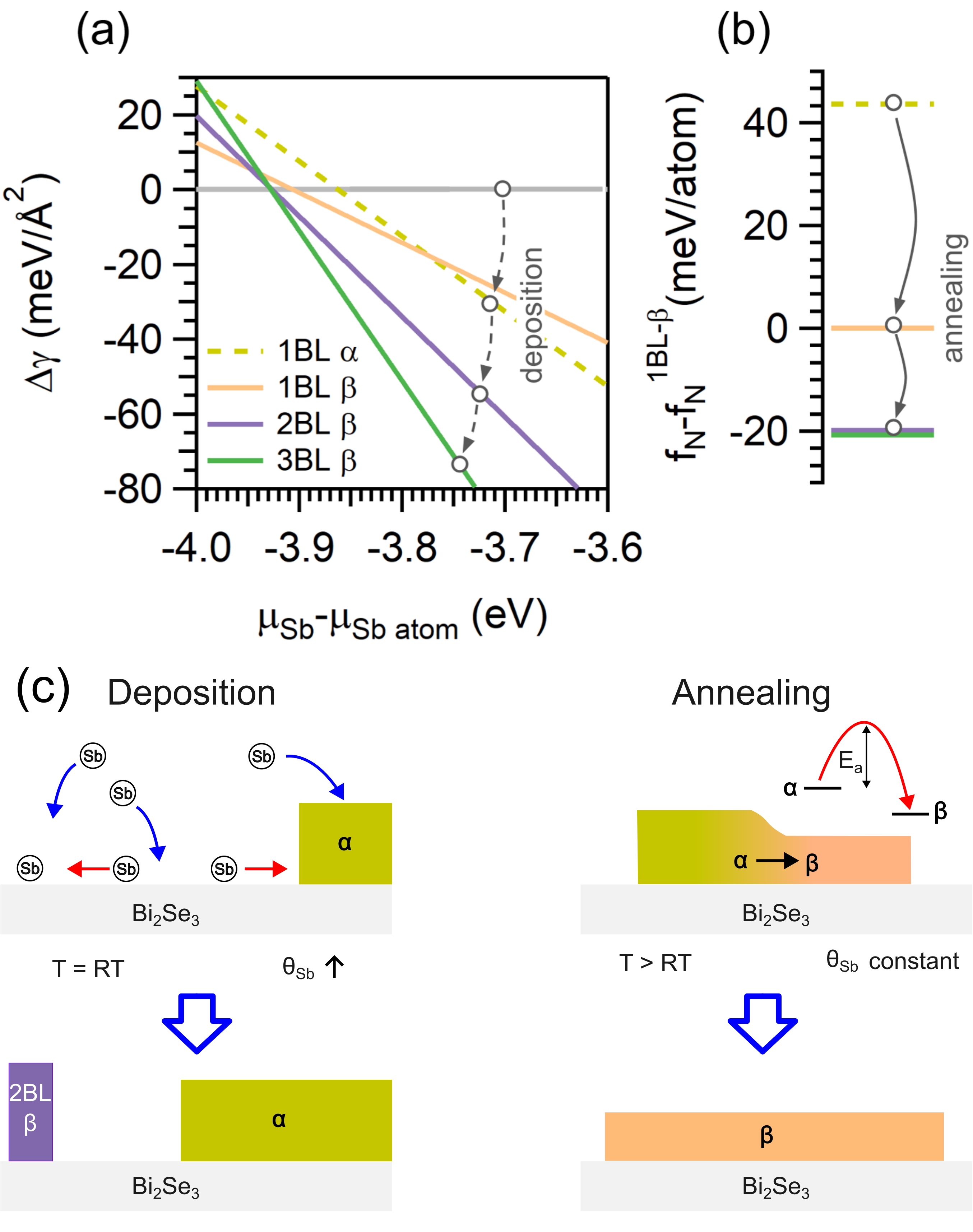}
	\caption{
(a) Thermodynamic phase diagram for $\alpha$ and $\beta$ phases of antimonene on Bi$_2$Se$_3$ in the presence of the Sb flux. As deposition continues, phases of higher coverage become thermodynamically accessible (dashed arrows). (b) Formation energy per atom during the annealing stage, in the absence of the gaseous reservoir. Possible phase transformations are indicated by solid arrows. (c) Schematic depiction of the deposition and annealing mechanisms.}  \label{Fig-energy}
\end{figure}

In order to explain the observed growth processes, we assessed the relative stability of the various Sb/\BiSe\ phases using atomistic thermodynamics and DFT.  
Fig.~\ref{Fig-energy}(a) shows the formation energy per area ($\Delta\gamma$) computed from the grand (Landau) potential \cite{HoganACSNano2019}. It describes the relative stability of a particular adlayer during a deposition process in the presence of a gaseous flux, and depends on the Sb chemical potential ($\Delta \mu = \mu_{Sb}-\mu_{Sb~atom}$), DFT total energy, and surface atomic density. 
Instead, the formation energy per atom ($\Delta f_N$, displayed in Fig.~\ref{Fig-energy}(b) depends solely on the total energy and directly determines the favoured geometry during the annealing stages, when the coverage is kept fixed \cite{Hogan_PRB_pnictogen}. 
Although the computed energies refer to uniformly covered substrates and not finite islands, with kinetic processes not considered important, information can nonetheless be gleaned from the calculations.
In the context of the real experiments performed in this work, the phase diagram explains the geometry of a specific large island in terms of the local Sb coverage. A value of about $\Delta\mu=-3.7$~eV explains most of the observed data.

During the initial deposition stage, the average Sb coverage remains relatively low and therefore islands of no more than 1 BL thickness are expected to form. According to Fig.~\ref{Fig-energy}(a), growth of pure $\alpha$ (or mixed $\alpha$/$\beta$) domains are thermodynamically favoured for $\Delta\mu \ge -3.75$~eV. 
Indeed, the 1 BL $\alpha$ phase is always favoured over the 1 BL $\beta$ in the limit of large (positive) chemical potential for any pnictide adatom.\cite{Hogan_PRB_pnictogen}
This is consistent with the STM observations in Fig.~\ref{Fig-STM}(a) (note the absence of large 1BL $\beta$ domains) and the featureless ARPES spectrum (`1$\alpha$ islands' in Fig.~\ref{Fig-ARPES}(b)) appearing after the initial Sb deposition.
Nonetheless, on sample areas with a higher local Sb concentration, more stable phases of higher coverage may appear Fig.~\ref{Fig-energy}(c). Thus local island growth of 2--3 BL $\beta$ is observed in the top-right corner of Fig.~\ref{Fig-STM}(a).
With further Sb deposition across the substrate, 
thicker $\beta$ phases are predicted to form, as indeed detected along the horizontal path tracked by the ARPES measurements (uppermost cyan path in Fig.~\ref{Fig-ARPES}(b)). 

Once the flux is removed, the number of Sb atoms is kept constant and the stability is simply determined by $\Delta f_N$ as shown in Fig.~\ref{Fig-energy}(b). The as-deposited 1 BL $\alpha$ phase results to be less stable than all other phases under these new conditions. Depending on the kinetic energy barrier \cite{HoganACSNano2019,Hogan_PRB_pnictogen}, annealing time and temperature, island size and available space on the substrate, it can transform to a more stable configuration Fig.~\ref{Fig-energy}(c). 
In particular, the 1 BL $\alpha \to \beta$  transformation is clearly identified in Fig.~\ref{Fig-STM}(b) and along the vertical black/brown path in Fig.~\ref{Fig-ARPES} leading to `1$\beta$ islands'.
By recording STM images as a function of annealing time, we previously demonstrated \cite{HoganACSNano2019} that this indeed occurs
and estimated, using nudged-elastic-band calculations, a transition barrier of only ~240 meV/atom, i.e. quite accessible at the present experimental conditions of annealing temperature and time \cite{Hogan_PRB_pnictogen}.
With further annealing, large 1 BL $\beta$ islands are also predicted to transform to more stable (albeit smaller, via a dewetting mechanism) 2 or 3 BL $\beta$ islands, as indeed observed in Fig.~\ref{Fig-STM}(c) and Fig.~\ref{Fig-ARPES}(b) (vertical black path, terminating in `2$\beta$ islands').
The 1 BL $\alpha \to \beta$ annealing transition is also observed for mixed as-deposited phases (e.g. last step of cyan path). Finally, once all the $\alpha$ phase has been (irreversibly \cite{HoganACSNano2019}) removed via annealing, the phase diagram in Fig.~\ref{Fig-energy}(a) becomes valid again as further deposition implies, as before, the progressive growth of higher coverage phases on top of the 1~BL $\beta$ layer. This is clearly observed along the horizontal brown path in Fig.~\ref{Fig-ARPES}(b) passing through `2$\beta$ sheet' and terminating at `3$\beta$ sheet'. Note that for a three-step deposit-anneal-deposit procedure, the relative stability of the various Sb phases is in turn governed by $\Delta\gamma$, then $\Delta f_N$, and then again by $\Delta\gamma$: grown phases that suddenly appear metastable when the boundary conditions change may remain due to kinetic barriers.

%%%%%% CORE LEVEL %%%%%%%

\begin{figure*}[t]
	\includegraphics[width=1\textwidth]{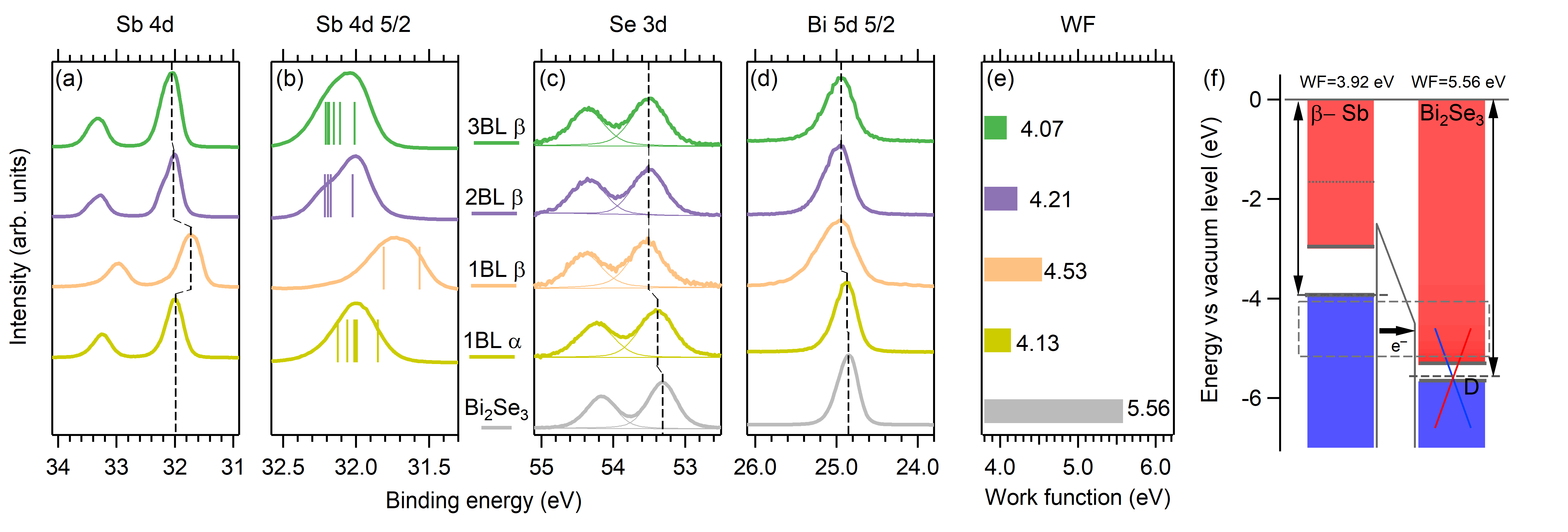}
	\caption{Core level spectra for (a,b) Sb\textit{4d}, (c) Se\textit{3d} and (d) Bi\textit{5d$_{5/2}$} relative to the bismuth selenide substrate and to the antimonene phases. The data are taken at RT with a photon energy of 75 eV. A linear background was subtracted from the spectra. Vertical bars under the Sb \textit{4d$_{5/2}$} spectra indicate the energy position of the calculated core level shifts. (e) The calculated work functions are represented by colored bars, with longer bars indicating higher values. (f) Band alignment between $\beta$ free-standing antimonene and Bi$_2$Se$_3$ substrate. Label D indicates the Dirac point of the \BiSe\ substrate.}  \label{Fig-CLS}
\end{figure*}

As mentioned above, the $\alpha$ phase does not show any clear spectroscopic signature in the valence states. For this reason we performed systematic core-level spectroscopy measurements in surface-sensitive conditions. The spectra shown below were acquired once the substrate was fully covered (sheet) and the antimonene phases were ascertained by ARPES. The data for Sb\textit{4d}, Se\textit{3d} and Bi\textit{5d} core levels are reported in Fig.~\ref{Fig-CLS}(a-d). The Sb\textit{4d} level relative to the $\alpha$ phase (Fig.~\ref{Fig-CLS}(a,b)) shows an approximate binding energy of 31.95 eV for the 5/2 peak (33.20 eV for the 3/2 peak), as extracted from the peak fitting procedure (with an estimated uncertainty of $\pm$ 0.05 eV). Upon transition to 1 BL $\beta$, a rigid shift of about 0.25 eV towards lower binding energies is recorded, with the emergence of a broader component at 31.70 eV. The shoulder at higher binding energies can be related to remnants of the 1 BL $\alpha$ phase, due to an incomplete phase transition.
Upon transition to 2 BL $\beta$, the relative core level experiences a backward shift of about 0.3 eV (32.00 eV), with a clear second component emerging at higher binding energies. The Sb\textit{4d} peak of the 3 BL $\beta$ structure displays nearly the same binding energy of the 2 BL $\beta$ case, though with a more pronounced second component. 

We also followed the behaviour of Bi\textit{5d} and Se\textit{3d} core levels and found that they show a continuous positive CLS from the growth of the $\alpha$ phase, to the transition to 1 BL $\beta$, up to thicker coverage (Fig.~\ref{Fig-CLS}(c,d)). The largest CLS occurs for the transition from 1 BL $\alpha$ to 1 BL $\beta$ with negative core level shift of about 0.10 eV for Bi\textit{5d} and 0.15 eV for Se\textit{3d} peaks. The large CLS of Sb\textit{4d} could be used to identify in particular the presence of the 1 BL $\beta$ allotrope in mixed phase samples by means of photoelectron spectroscopy studies. From the binding energy positions of Se\textit{3d}, Bi\textit{5d} and their relative intensities, one can also be sensitive to 1 BL $\alpha$, 2 BL and thicker $\beta$ phases.

However, the core level spectra do not show notable spectroscopic signatures, only the corresponding ARPES maps allowed us to make the assignment to a specific antimonene phase. Calculated binding energy shifts (coloured bars) have been overlaid on the Sb\textit{4d$_{5/2}$} core level spectra and displayed in Fig.~\ref{Fig-CLS}(b) to complement the experimental outcome. The bars relative to all spectra have been rigidly shifted by 31.85 eV to account for the measured binding energy of the $\alpha$ phase. We considered six non--equivalent atoms components for 1 BL $\alpha$ and two for 1 BL $\beta$. Four components are expected for 2 BL $\beta$, with three of them nearly degenerate in energy, and six for 3 BL $\beta$. A negative CLS is predicted for 1 BL $\beta$ with respect to 1 BL $\alpha$, and positive shift for 2 BL $\beta$ with respect to the 1 BL $\beta$ allotrope. The calculations show an overall good agreement, with a negative CLS for 1 BL $\beta$ and similar binding energies for the other phases. 

In order to understand more deeply the binding energy shifts in the photoemission spectra, we calculated the work functions (WFs) for different antimonene allotropes on the substrate, as well as for the clean substrate alone (Fig.~\ref{Fig-CLS}(e)). All the allotropes show a significant decrease of the WF with respect to Bi$_2$Se$_3$, in line with their metallic character. The change of the WF mimics the behaviour of the Sb\textit{4d} BE, with the highest WF value (of 4.53 eV) exhibited by the 1 BL $\beta$. We notice that the WFs values differ at least by 0.05 eV among the different phases; therefore, different allotropes can be distinguished by careful measurement of secondary electrons cutoff, for instance by photoemission microscopy or by Kelvin probe techniques.

By using the above WF values, we deduced the band alignment between Bi$_2$Se$_3$ and free-standing 1 BL $\beta$-antimonene (WF=3.92 eV), as shown in Fig.~\ref{Fig-CLS}(f). The valence band maximum (VBM) of antimonene lies well above the conduction band minimum (CBM) of Bi$_2$Se$_3$, resulting in a so-called type-III (or broken gap) alignment \cite{hu2017two}. 
Despite the expected gap underestimation (0.93 eV), typical of PBE calculations, for antimonene we expect the same band alignment for an electronic gap up to 2.30 eV, as marked by dotted lines within the CB and VB of free-standing antimonene. 
As a result, a strong transfer of electrons to Bi$_2$Se$_3$ occurs, so that 1 BL $\beta$--antimonene becomes $p$-doped while the Bi$_2$Se$_3$, $n$-doped. This can also be confirmed by the core level shifts of Se and Bi (Fig.~\ref{Fig-CLS}(c,d)). A strong perpendicular dipole is hence created at the interface between Sb and topmost Se atoms. For thicker coverage, the surface field gradient reduces gradually leading to a Sb\textit{4d} core level BE, close to its bulk value. The dipole indeed remains buried and possibly neutralized at the interface, as demonstrated by the Se and Bi core levels that barely change their BEs.

We emphasise that the Sb\textit{4d} value of 31.70 eV BE, for 1 BL $\beta$-antimonene, is the lowest value reported for similar systems. Indeed, on substrates such as PdTe$_2$, Bi$_2$Te$_3$ and Sb$_2$Te$_3$, the Sb\textit{4d} core level was found in the 32.0--32.3 eV BE range \cite{TaoLei_JAP_2016, su2021topological, WuADMA2017}. As a consequence, the WFs values approach that of the free-standing $\beta$--antimonene, leading to much less charge transfer at the interface. This is why the position of the Dirac point of the migrated TSS in those systems, is located very close to the Fermi level \cite{JinPRB2016, TaoLei_JAP_2016, su2021topological}. 

More generally, due to the buckled structure of 2-dimensional materials, the value of the electronic gap, the work function, and hence the amount of the charge transfer, might be tuned by strain \cite{ShiNL2020,Ji_JPCC2022_alfa_beta}. Doping is also another way to change the electronic properties \cite{NGUYEN2022115315_doping}. A notable example is bismuthene, for which a large WF difference and therefore a broken gap band alignment is also predicted, leading to a strongly \textit{p-}doped Dirac cone similar to antimonene \cite{JinPRB2016, su2017selective}. Such kind of band alignment allows for an increase of the tunnelling current density \cite{ozccelik2016band}, possibly leading to the negative differential resistance phenomenon, of paramount importance for tunnelling field effect transistors. 

\section{Conclusions}\label{sec:conclusion}

A detailed study on the growth of antimonene allotropes on Bi$_2$Se$_3$ was conducted using STM, ARPES, and core-level photoemission, complemented by DFT based electronic band structure and thermodinamics calculations. Growth protocols were identified to form single-phase allotropes that can be epitaxially grown on Bi$_2$Se$_3$. The puckered $\alpha$ phase forms at room temperature but becomes unstable beyond full coverage or at higher temperatures, where it transforms into the more stable hexagonal $\beta$ allotrope. Large-scale, single-phase monolayer (1 BL) $\beta$ structures emerge only after annealing, necessitating a two-step process (growth + annealing). In contrast, achieving a complete single phase of 2 BL $\beta$ or thicker depends on using the 1 BL $\beta$ structure as a template, necessitating a three-step process (growth + annealing + growth). The 1 BL $\beta$-antimonene exhibits an unusually high Sb 4$d$ core-level shift due to the formation of an interface dipole, which is compensated in thicker layers. This dipole leads to a type-III band alignment, potentially useful in tunneling field-effect transistors.

\section*{Acknowledgements}
The technical assistance of G. Emma, G. De Santis, L. Sancin and F. Zuccaro is greatly acknowledged. We thank Elettra Sincrotrone Trieste for providing access to its synchrotron radiation facilities (proposal nr. 20195461 and 20215019). C.C, P.M. and P.M.S. acknowledge EUROFEL-ROADMAP ESFRI of the Italian Ministry of University and Research. S.V.E. acknowledges support from the Government research assignment for ISPMS SB RAS, project FWRW-2022-0001.
E.V.C. acknowledges support from Saint Petersburg State University (Project ID No. 95442847). C.H. acknowledges the CINECA award under the ISCRA initiative, for the availability of high performance computing resources and support. Z.R.B. and S.G. acknowledge financial support from the Slovenian Research Agency (research core funding no. P2-0412).

\section*{Data Availability Statement}
The data that support the findings of this study are available from the corresponding author upon reasonable request.

%

%\bibliographystyle{apsrev4-1}
%\bibliography{Biblio-fingerprints_000.bib}

%\begin{thebibliography}{51}%
%\end{thebibliography}%

\end{document}